\documentclass{caosp}
\usepackage{graphicx}

\articleNo{193}
\pubyear{2010}
\volume{40}
\volnumber{1}
\firstpage{1}
\received{December 23, 2009}
\accepted{March 29, 2010}

\begin{document}

%%%%%%%%%%%%%%%%%%%%%%%%%%%%%%%%%%%%%%%%%%%%%%%%%%%%%%%%%%%%%%%%%%%%%%%%%%%%%
%              R U N N I N G   P A G E   H E A D I N G S
% Odd page headings (except for the title page) are produced automatically
% and contain the title. If, and only if, the title of your article is too
% long the running head is ommitted in the printout; you can make your own
% running title by using the \htitle command, putting the shortened title
% between the curly brackets. \htitle should also be used when the
% subtitle is present: \htitle offers you a way how to include it into the
% headings. If you wish to see how it works simply remove the % sign from
% the beginnig of that line.
%
% Unlike the \htitle command, the \hauthor command is compulsory. It is
% used to produce even page headings and contains the names of the authors
% of an article.  All authors must be listed here, if possible. When
% authors' list is too long, you can abbreviate it by using "{\it et
% al.}". Authors' names are given in the form: initial(s) of the author's
% first name and surname. Authors are separated by a "," (comma) sign and
% the last one by "and".
%%%%%%%%%%%%%%%%%%%%%%%%%%%%%%%%%%%%%%%%%%%%%%%%%%%%%%%%%%%%%%%%%%%%%%%%%%%%%
\htitle{The SECIS instrument on the Lomnick\'{y} Peak Observatory}
%\hauthor{J.\,Ambr\'{o}z {\it et al.}}
\hauthor{J.~Ambr\'{o}z, K.~Radziszewski, P.~Rudawy, J.~Ryb\'{a}k,
         K.J.H.~Phillips}

%%%%%%%%%%%%%%%%%%%%%%%%%%%%%%%%%%%%%%%%%%%%%%%%%%%%%%%%%%%%%%%%%%%%%%%%%%%%%
%                       T I T L E
% Capital letters in the title are only used at the beginning of the
% names. Don`t end the title by a "." (dot)
%%%%%%%%%%%%%%%%%%%%%%%%%%%%%%%%%%%%%%%%%%%%%%%%%%%%%%%%%%%%%%%%%%%%%%%%%%%%%
\title{The SECIS instrument on the Lomnick\'{y} Peak Observatory}

%%%%%%%%%%%%%%%%%%%%%%%%%%%%%%%%%%%%%%%%%%%%%%%%%%%%%%%%%%%%%%%%%%%%%%%%%%%%%
%                       S U B T I T L E
% You can use the subtitle, with the command \subtitle similar to the
% \title command.
%%%%%%%%%%%%%%%%%%%%%%%%%%%%%%%%%%%%%%%%%%%%%%%%%%%%%%%%%%%%%%%%%%%%%%%%%%%%%

%%%%%%%%%%%%%%%%%%%%%%%%%%%%%%%%%%%%%%%%%%%%%%%%%%%%%%%%%%%%%%%%%%%%%%%%%%%%%
%                   A U T H O R  N A M E S
% Authors' names are separated by the \and commmand and their institutes
% are assigned by the \inst{n} command.
%
% When the name contains "Slovak" letters L,d,t,l with a caron, use an
% \additional softl, etc. command (examples given in the last table of
% this document) to produce typographically correct accented characters.
%%%%%%%%%%%%%%%%%%%%%%%%%%%%%%%%%%%%%%%%%%%%%%%%%%%%%%%%%%%%%%%%%%%%%%%%%%%%%
\author{
        J.~Ambr\'{o}z \inst{1}
      \and
        K.~Radziszewski \inst{2}
      \and
        P.~Rudawy \inst{2}
      \and
        J.~Ryb\'{a}k \inst{1}
      \and
        K.~J.~H.~Phillips \inst{3}
       }

%%%%%%%%%%%%%%%%%%%%%%%%%%%%%%%%%%%%%%%%%%%%%%%%%%%%%%%%%%%%%%%%%%%%%%%%%%%%%
%           I N S T I T U T E S'  A D D R E S S E S
% The affiliation of authors is generated by the \institute command, the
% \and command being again used to separate individual addresses.
% The following commands may be used for the following three institutes:
%               \lomnica        for      AsU SAV, Tatranska Lomnica
%               \blava          for      AsU SAV, Bratislava
%               \ondrejov       for      AsU CAV, Ondrejov
%
% The given postal address must be complete in order to facilitate our
% editorial work. Moreover, you can add your e-mail address, using the
% \email command.
%%%%%%%%%%%%%%%%%%%%%%%%%%%%%%%%%%%%%%%%%%%%%%%%%%%%%%%%%%%%%%%%%%%%%%%%%%%%%
\institute{
           \lomnica, \email{ambroz@astro.sk}
         \and
	 Astronomical Institute of Wroc{\l}aw University,\\ 51-622 Wroc{\l}aw,
	 ul. Kopernika 11, Poland
         \and
	 University College London -- Mullard Space Science Laboratory,
         Holmbury St Mary,\\ Dorking, Surrey RH5 6NT, UK
          }

%%%%%%%%%%%%%%%%%%%%%%%%%%%%%%%%%%%%%%%%%%%%%%%%%%%%%%%%%%%%%%%%%%%%%%%%%%%%%
%                        D A T E / R E C E I V E D
% Date inserted here will be the date when your paper was received The
% format is: month (not abbreviated), day, year.
%%%%%%%%%%%%%%%%%%%%%%%%%%%%%%%%%%%%%%%%%%%%%%%%%%%%%%%%%%%%%%%%%%%%%%%%%%%%%
\date{March 29, 2003}

%%%%%%%%%%%%%%%%%%%%%%%%%%%%%%%%%%%%%%%%%%%%%%%%%%%%%%%%%%%%%%%%%%%%%%%%%%%%%
%                        M A K E T I T L E
% The beginning part (title, author(s), etc.) of your article must be
% closed by the \maketitle command.
%%%%%%%%%%%%%%%%%%%%%%%%%%%%%%%%%%%%%%%%%%%%%%%%%%%%%%%%%%%%%%%%%%%%%%%%%%%%%
\maketitle

%%%%%%%%%%%%%%%%%%%%%%%%%%%%%%%%%%%%%%%%%%%%%%%%%%%%%%%%%%%%%%%%%%%%%%%%%%%%%
%                        A B S T R A C T,  K E Y W O R D S
% Here it is shown how to write an abstract.  Keywords should be placed
% within the "abstract" environment using the command \keywords and they
% should be selected from the thesaurus from Astron.  Astrophys.
% Abstracts. They must be separated from each other by -- (two dashes).
%%%%%%%%%%%%%%%%%%%%%%%%%%%%%%%%%%%%%%%%%%%%%%%%%%%%%%%%%%%%%%%%%%%%%%%%%%%%%
\begin{abstract}
Heating mechanisms of the solar corona will be investigated at the
high-altitude solar observatory Lomnicky Peak of the Astronomical Institute
of SAS (Slovakia) using its mid-size Lyot coronagraph and post-focal
instrument SECIS provided by Astronomical Institute of the
University of Wroc{\l}aw (Poland).
The data will be studied with respect to the energy
transport and release responsible for heating the solar corona
to temperatures of mega-Kelvins. In particular investigations
will be focused on detection of possible high-frequency MHD waves
in the solar corona.
The scientific background of the project, technical  details of the SECIS
system modified specially for the Lomnicky Peak coronagraph,
and inspection of the test data are described in the paper.
\end{abstract}

%%%%%%%%%%%%%%%%%%%%%%%%%%%%%%%%%%%%%%%%%%%%%%%%%%%%%%%%%%%%%%%%%%%%%%%%%%%%%
%                       S E C T I O N I N G
% Any section starts with the command \section as shown below, with the
% title in Initial Capitals and lowercase only. Do not number the sections
% - let LaTeX do that for you - and do not end them by a "." (dot).
%
% The (sub)section titles are typeset in boldface; so, if working in the
% mathematics mode in (sub)section titles, you must use \boldmath and
% enclose it into curly brackets, e.g. "{\bolmath $R^{2}$}".
%%%%%%%%%%%%%%%%%%%%%%%%%%%%%%%%%%%%%%%%%%%%%%%%%%%%%%%%%%%%%%%%%%%%%%%%%%%%%
%Section 1
\section{Introduction}
%%%%%%%%%%%%%%%%%%%%%%%%%%%%%%%%%%%%%%%%%%%%%%%%%%%%%%%%%%%%%%%%%%%%%%%%%%%%%
%                       L A B E L
% The label command is very convenient for you when referring to sections,
% subsections,..., tables, figures as well as to equations (see commands
% \ref and \pageref). In the case of figure and/or table environments the
% \label command should always be put after the \caption command to
% preserve proper numbering. When using the \label command the file must
% be compiled twice to get proper cross-references.
%%%%%%%%%%%%%%%%%%%%%%%%%%%%%%%%%%%%%%%%%%%%%%%%%%%%%%%%%%%%%%%%%%%%%%%%%%%%%
%\label{introduction}
For more than 60 years it has been well known that the quiet solar corona
is heated to a temperature of about 1--2 million
Kelvins while the visible surface of the Sun is roughly 250 times cooler
(Grotrian 1939; Edlen 1942; Phillips, 1995). It has been also recognized
that magnetic fields or waves play a key r\^{o}le in the heating of the solar
corona so that somehow convective energy in the photosphere is converted
to thermal energy in the corona via magnetic
fields or wave energy. The primary energy source for this heating
must lie in the
convection zone below the solar photosphere
(e.g. Bray et al., 1991; Golub \& Pasachoff 1998; Aschwanden 2004)
where there is 100 times as much energy available than that required to heat
the corona ($\approx$300\,W/m$^2$: Withbroe \& Noyes 1977, Aschwanden 2001).

Currently, the debate centres on whether the energy to heat the corona
derives from dissipation of magneto-hydrodynamic (MHD) waves
(e.g. Hollweg, 1981) or from numerous small-scale magnetic
reconnection events giving rise to nanoflares
(Parker, 1988, Aschwanden 2004). It has been found theoretically that
the interaction of the magnetic field
with convective flows in or below the photosphere can produce two
types of magnetic disturbances in coronal structures. Firstly, the
buffeting of
magnetic flux concentrations in the photosphere by granulation
generates MHD waves which can propagate
into magnetic flux tubes and dissipate their energy in the
chromosphere or corona (e.g. Ofman et al. 1998). Secondly, in coronal
loops the random motions of magnetic loop foot-points can produce twisting
and
braiding of coronal field lines, which generates field-aligned electric
currents that can be dissipated resistively (e.g. Parker 1972, 1983;
Heyvaerts \& Priest 1983; van Ballegooijen 1990). The main difference
between these processes is that plasma inertia plays a key r\^{o}le in wave
propagation, but is unimportant for the dynamics of field-aligned
currents along coronal loops. Thus these types of
magnetic heating mechanisms can be crudely classified as either
wave-heating or current-heating mechanisms.

There are theoretical arguments for both mechanisms, but the
observational evidence for nano-flare heating is perhaps looking
 less convincing than before. Extrapolation of the number spectra
of small flares down to microflares has been made to nano-flares
but the total energy, while tantalisingly close, is most probably
less than the required amount (Parnell et al. 2000).

%During the last decade many new results have been presented describing the
%behaviour of observable quantities like spectral line emissivity, Doppler shifts,
%and broadening of coronal and transition region lines (e.g. Hansteen 1993,
%Wikstol et al. 2000a, 2000b, Muller et al. 2004, Bogdan et al. 2003, Taroyan et al.,
%2006). These workers used 1D or 2D numerical
%(magneto)-hydrodynamic codes with different boundary conditions, the results
%showing that different heating mechanisms can give rise to different observable
%quantities, though at present observations lack the resolution to make a proper
%distinction. It is nonetheless  an urgently needed task if the coronal heating
%problem is to be solved (Klimchuk 2006).

%%%%%%%%%%%%%%%%%%%%%%%%%%%%%%%%%%%%%%%%%%%%%%%%%%%%%%%%%%%%%%%%%%%%%%%%%%%%%
%Section 2
\section{High-frequency loop oscillations}
%\label{oscillations}
Several theoretical
studies showed that only high-frequency MHD waves ($>1$\,Hz) are capable
of significant heating (e.g. Porter et al. 1994, Aschwanden 2004).
These waves have been sought using the Fe XIV ``green" line at 530.3\,nm
(emitted at $\sim 2\times10^6$~K) and the Fe X ``red" line at 637.5~nm
($\sim 1.2\times 10^6$~K). Observations of high-frequency intensity
oscillations of the coronal structures have been made
by Pasachoff and colleagues
(green line: Pasachoff et al., 1995, 2000, 2002),
by Ru\v{s}in and Minarovjech
(green and red lines: Ru\v{s}in \& Minarovjech, 1991, 1994),
by Rudawy, Phillips and colleagues
(green line, total eclipses in 1999 and 2001;
Phillips et al., 2000, Williams et al., 2001;
Rudawy et al., 2001; Williams et al., 2002; Rudawy et al., 2004), and by
Singh et al.
(green and red lines: Singh et al., 2009).
Phillips and Rudawy and their colleagues with their SECIS (Solar Eclipse
Coronal Imaging System) CCD camera instrument have obtained the highest time
resolution up to now. The results of these investigations are somewhat
contradictory, with both positive and negative observations of oscillations
(e.g. Pasachoff \& Landman 1984; Koutchmy et al. 1994;
Cowsik et al. 1999; Williams et al. 2001, 2002; Rudawy et al. 2004).
%
%Our project extends studies already done by Phillips, Rudawy, and
%co-workers in a joint British--Polish project designed to investigate
%high-frequency (sub-second) brightness oscillations in the corona with
%the SECIS (Solar Eclipse Coronal Imaging System) instrument.
Space missions capable of comparable time resolution measurements have not
been available up to the present time, so using ground-based equipment is still
the only way of making such observations (Aschwanden, 2004, Klimchuk, 2006).

In this paper, we describe a set-up that will be used to search for
high-frequency coronal oscillations.
We shall be making observations using the mid-sized coronagraph (belonging
to the Astronomical Institute of Slovak Academy of Sciences in Tatranska
Lomnica: Lexa, 1963) at Lomnicky Peak Observatory.
%, located at the summit of the Lomnicky Peak (2634\,m altitude)
Observations will be made in the
Fe~XIV green line using the SECIS instrument (Phillips et al. 2000), now
owned by the Astronomical Institute at the University of Wroc{\l}aw, Poland.

%Local intensity changes in various coronal structures will be recorded
%with high time resolution ($<$1/10 s).
%Similar variations of the intensity, already observed by the Anglo-Polish
%team during total solar eclipses in 1999 and 2001, may be caused
%by the dissipation of energy contained in high-frequency MHD waves
%in the corona generated in active regions as well as in some
%individual solar structures like loops and streamers. The dissipated
%energy can make a significant contribution to the total energy
%budget of the solar corona, but the exact amount of energy involved
%in MHD waves is still unknown.

%Measurements will be coordinated with other ground-based observatories
%and with space-borne instruments, in particular the ultraviolet
%instruments on {\it SOHO} (CDS and EIT), {\it TRACE}, the TESIS instrument on
%{\it CORONAS-PHOTON}, {\it STEREO}, and the EIS and XRT instruments
%on {\it Hinode}.

%%%%%%%%%%%%%%%%%%%%%%%%%%%%%%%%%%%%%%%%%%%%%%%%%%%%%%%%%%%%%%%%%%%%%%%%%%%%%
%Section 3
\section{Instrumentation}
%\label{Instrumentation}
The observational system on the Lomnicky Peak Observatory consists of 3
main instrumental parts: the 20-cm Lyot-type coronagraph, SECIS
instrument (two fast-frame-rate
CCD cameras, auxiliary electronics systems and
dedicated computer) and a special opto-mechanical interface
between the coronagraph and SECIS cameras.

%Subsection 3.1
\subsection{Coronagraph}
%\label{Coronagraph}
The Lomnicky Peak coronagraph (Lexa, 1963), made by Carl Zeiss Jena,
is located at the summit of Lomnicky Peak (2634~m altitude), allowing
observations in the light of prominent coronal visible-light emission lines
out to a significant distance beyond
the solar limb.

The front part of its optical system consists of a single objective
lens (BK7 glass, $R_1$=1.71\,m, $R_2$=17.0\,m, $D=$\,200\,mm, $f=$\,3\,m)
and a primary diaphragm which obscures the lens to a final
clear aperture of 195\,mm. The focal lengths of the objective for the
wavelength of the green line are 2975 and 2980\,mm for the
axial and outer light beams respectively.

The central part of the optical system has an artificial Moon (the occulting
disk) which is a fat mirror inclined with respect to the optical axis and
reflects the solar disk light out of the coronagraph tube. The artificial Moon
is fixed in front of a field lens in a hole in the center of the lens,
and can be changed to similar ones with various diameters.
Behind the field lens there is a re-imaging triplet lens in
order to correct, at least partially, geometric aberrations of the primary
lens and to focus a difraction image of the primary diaphragm on Lyot's stop.
Lyot's stop lies between the second and the third lenses of the triplet and
blocks the scattered light coming from the primary diaphragm.
A particular feature of the optical system is
a four-lens imaging objective of 9\,cm aperture. The combined
action of both objectives creates the final image of the corona
with diameter $\approx 40$\,mm.

The coronagraph is equipped with a fast optical
automatic guider. It detects offsets of the actual pointing using two
photodiode pairs in an anti-parallel connection providing
closed-loop correction signals to the drives. It precisely stabilizes the
relative position of the occulting disk against the solar image,
ensuring stable position of the field of view.

%Fig. 1
\begin{figure}
\centerline{\includegraphics[width=9.5cm,clip=]{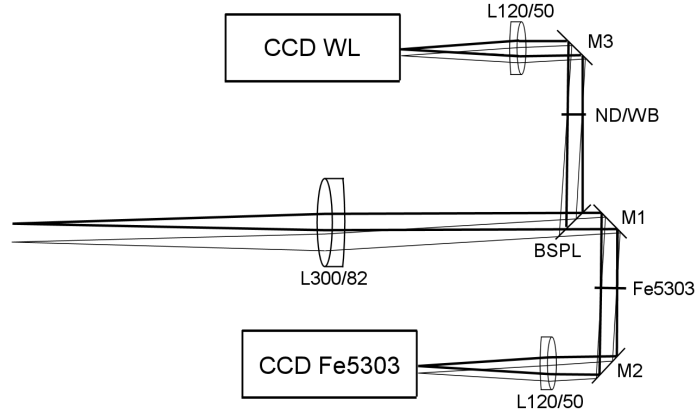}}
\caption{Scheme of the opto-mechanical interface connecting the Lomnicky
Peak coronagraph and the SECIS instrument.
The entrance lens L300/82 forms a collimated beam which is split by the
beam splitter (BSPL) into two beams, which are brought to a focus on the
two cameras (CCD WL, CCD Fe5303) by achromatic lenses L120/50 via the
narrow-band filter (Fe5303) and neutral density/broad-band filter
combination (ND/WB). See text for remaining parts. An on-axis pair
of rays is shown by the thick lines, and a pair of rays from one
extreme of the coronagraph image is shown by thin lines.
}
%\label{fig-1-setup-sketch}
\end{figure}

%Subsection 3.2
\subsection{Opto-mechanical interface}
%\label{interface}
The two SECIS cameras are connected to the Lomnicky
Peak coronagraph using a new, special opto-mechanical interface.

%The optical set-up is shown schematically in Fig.~\ref{fig-1-setup-sketch},
The optical set-up is shown schematically in Fig.\,1,
%with Fig.~\ref{fig-2-setup-real} showing the components in the rigid
with Fig.\,2 showing the components in the rigid
light-tight box. The coronal image formed by the coronagraph is to the
%left in Fig.~\ref{fig-1-setup-sketch}, at the focus of the entrance
left in Fig.\,1, at the focus of the entrance
%lens (marked L300/82 in Fig.~\ref{fig-1-setup-sketch}: diameter and
lens (marked L300/82 in Fig.\,1: diameter and
focal length are D=82~mm, f=300~mm). This lens forms a parallel
light-beam which then passes to a beam splitter (BSPL, shorter
dimension = 50~mm). The reflected beam from the beam splitter then
passes through a broad-band and neutral density filter combination
(marked ``ND/WB filter" in the figure), is reflected again from a
flat mirror (M3), and finally is brought to a focus on to the CCD
camera (marked ``CCD WL") by an achromatic lens (L120/50).
The parallel beam transmitted by the beam splitter is reflected
by a flat mirror (M1), passes a narrow-band FeXIV 530.3\,nm interference
filter (``Fe5303"), is reflected by a second flat mirror (M2), then
brought to a focus on to the CCD camera (marked ``CCD Fe5303") by the
%acromatic lens (L120/50). Fig.~\ref{fig-1-setup-sketch} shows the ray
acromatic lens (L120/50). Fig.\,1 shows the ray
diagram for the configuration. In this figure, an on-axis pair of rays
is shown by thick lines. The thin lines represent rays from one extreme
of the coronagraph image. The angle between them (greatly exaggerated
in the figure) is no more than about 0.5 degrees. With such a small
angle, a negligible wavelength shift is produced by the interference
filter (Fe5303).

%Fig. 2
\begin{figure}
\centerline{\includegraphics[width=10.5cm,clip=]{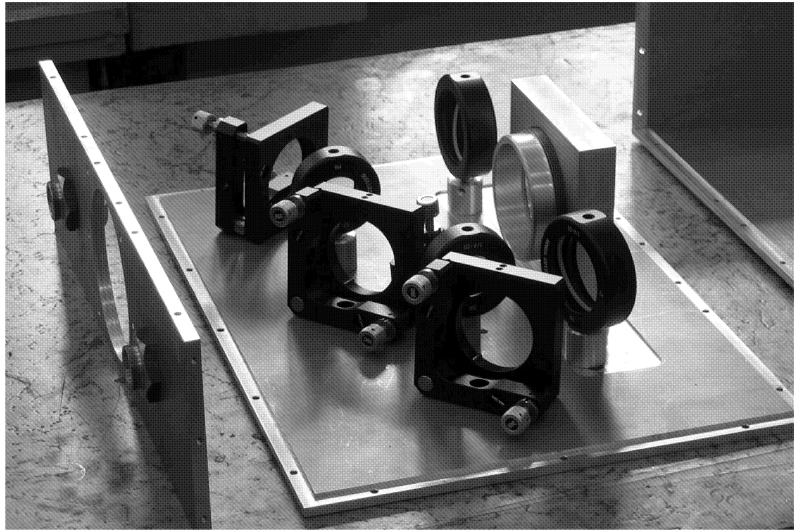}}
\caption{Rigid light-tight box acting as a frame for the opto-mechanical
interface before assembling to the final configuration.}
%\label{fig-2-setup-real}
\end{figure}

%Fig. 3
\begin{figure}
\centerline{\includegraphics[width=10.5cm,clip=]{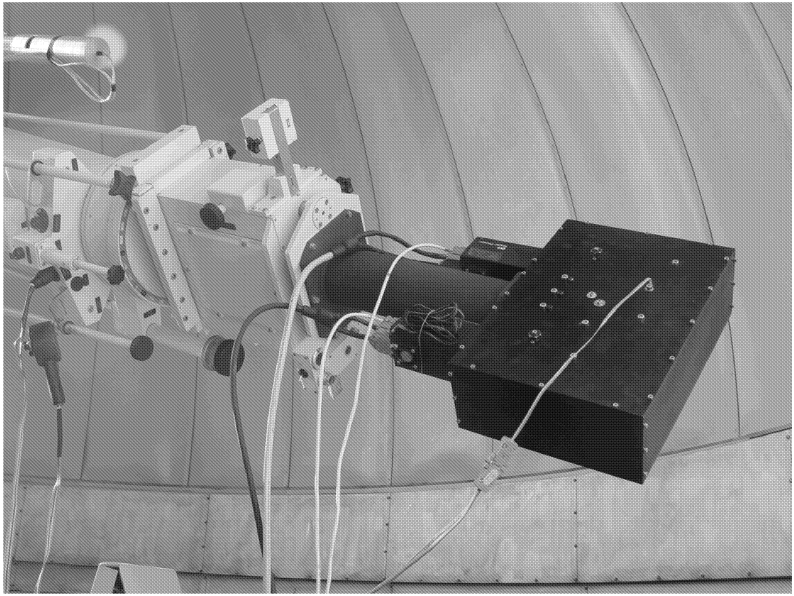}}
\caption{The opto-mechanical interface box in the test configuration
attached to the Lyot coronagraph at the Lomnicky Peak observatory
during tests in April 2009.}
%\label{fig-3-box-at-telescope}
\end{figure}

The optical system was designed taking into account the actual
optical parameters of the Lomnicky Peak coronagraph and the desired
spatial scale of the images on the CCD chips. It was optimized
to avoid any vignetting (to keep entire fields of view as bright
as possible) and to limit geometric or chromatic aberrations of
the system. In order to minimize the total cost of the interface,
all optical and mechanical elements as well as adjustable optical
mounts were general-purpose stock elements, selected from the Melles
Griot company catalogue.

The beam splitter (BSPL) reflects only 10\% of the
light to the broad-band channel, transmitting the remaining light to
the narrow-band (green-line)
channel. The  broad-band filter was selected to be centred on
the green line wavelength but having a larger range of transmission
(central wavelength 530.0\,nm, bandpass FWHM 10\,nm).
A neutral-density filter (ND) in the white-light optical channel was
selected to ensure equal exposure times in the narrow-band
and broad-band channels.
Final images in both channels are formed by lenses with the same
focal lengths, so giving the same spatial scale.

The entire optical system is mounted in a rigid box attached to a
rear plate of the coronagraph. A light-tight black aluminium box
acting as a rigid optical frame for the optical components was
manufactured by the workshop of the Astronomical Institute at the
%University of Wroc{\l}aw (Fig.~\ref{fig-2-setup-real}).
University of Wroc{\l}aw (Fig.\,2).
The optical system was assembled and pre-aligned in the box; apart
from focusing of the system, no other special alignments
of the optical components are needed at the telescope
%(Fig.~\ref{fig-3-box-at-telescope}).
(Fig.\,3).

%Subsection 3.3
\subsection{Narrow-band filter}
%\label{NBfilter}

Three narrow-band filters with passbands around the
FeXIV 530.3\,nm green line are available (two made by
Barr Associates, Inc., and one by Andover Corporation).
All filters have FWHM passbands of $\approx 0.25$~nm and
diameters of 50~mm. The filter chosen for use in the optical system
is fitted with a thermostatic device to maintain the required
temperature under variable ambient temperatures.

%Fig. 4
\begin{figure}
\centerline{\includegraphics[width=9.5cm,clip=]{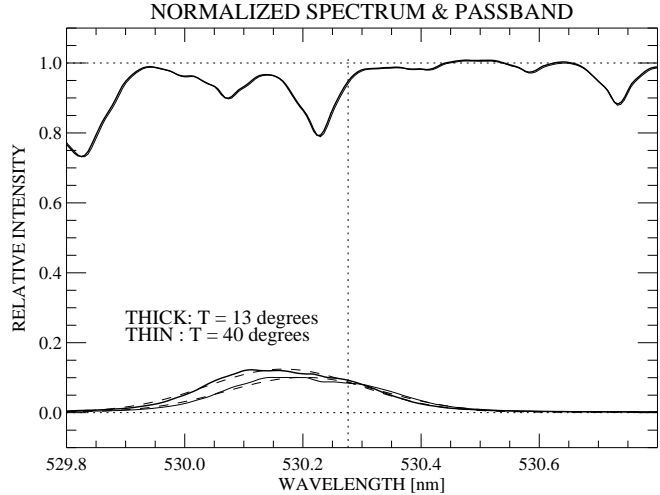}}
\caption{Reference disk centre spectrum taken without
the narrow-band filter using a small spectrograph (upper solid thick
line) and two examples of the transmission curves of the narrow-band
filter at temperatures of 13 and 40 degrees Celsius (bottom
thick and thin lines respectively). The dashed lines show Gaussian
fits of the measured transmission curves.
The vertical dotted line indicates the wavelength of the coronal green line.
}
%\label{fig-4-NB-filter-tests}
\end{figure}

%Fig. 5
\begin{figure}
\centerline{\includegraphics[width=9.5cm,clip=]{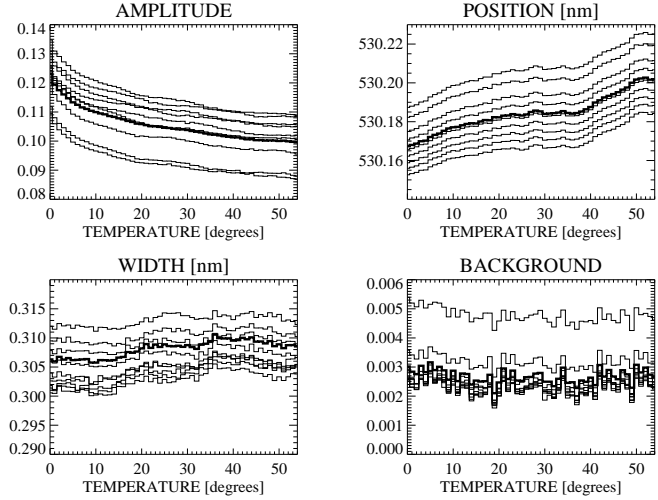}}
\caption{Dependence of the main passband parameters on the filter temperature
in the range 13--40 degrees Celsius
and on position across the filter surface:
maximum transmission (upper left panel), central wavelength (upper right),
FWHM width (bottom left), and background level (bottom right).
Intensities are normalized to the incoming light level.
%Thick curves refer to averaged results over a radial line with length
%7.2\,mm (360\,arcsecs) and
Thick curves refer to averages along a radial direction from the
centre of the filter over a length of 7.2mm (360\,arcsecs) and
thin curves to narrow regions along the radial direction (width 0.8~mm or
40~arcsecs).
%A typical maximum transmission of the filter is $\approx$10\%,
%the FWHM is $\approx 0.3$~nm for the wavelength of the green line.
%The individual results show a similar temperature response to those of the
%averaged results.
}
%\label{fig-5-NB-tests}
\end{figure}

%Subsection 3.4
\subsection{SECIS instrument}
%\label{SECIS}

The SECIS instrument
was built, tested, and calibrated between 1997 and 1999 in a British--Polish
collaboration to search for short-period
coronal light oscillations. It was used with great success during the total
eclipse seen from Bulgaria in 1999, from Zambia in 2001 (see
Phillips et al., 2000; Williams et al., 2001; Williams et al., 2002;
Rudawy et al., 2004), and most recently from Libya in 2006.

The CCD cameras (manufactured by EEV, Chelmsford, U.K.) are
high-performance cameras designed specifically for scientific
and machine vision applications. The image sensor is a monochrome
512$\times$512 pixel frame transfer CCD. This device has square
(15$\mu$m~$\times  15 \mu$m) pixels, and can be driven at a non-interlaced
frame rate up to 70 frames per second. The cameras digitise the signal
from the CCD to nominally 12 bits and provide a real dynamic range
of over 1000:1. The camera electronics operate the cameras in an
``asynchronous" mode, where a trigger pulse from the control
electronics commands one of the cameras to capture an image at
a precise moment and with a precise exposure period.
This feature allows the two cameras to capture accurately
synchronised images. The data are captured and stored on a personal computer.

The computer system captures the synchronised digital video streams from the
two CCD cameras and reconstitutes the video images, storing them
for more detailed analysis. The computer has
dual Pentium processors, 128 megabytes of memory, and four 9\,GB
disk drives. It is able to run a set of observations consisting of
up to about 10000 images for each camera. The image processing software
allows the replay of the video, and the cropping to sub-sequences
and regions of interest. These selections can then be exported to files in
FITS format.
More detailed information
about SECIS and its first scientific application are described
by Phillips et al. (2000).

Since 2003 SECIS has also been used for making
high time resolution spectral observations
of solar flares over the profile of the H-$\alpha$ (656.3\,nm) line
using the Multi-Channel Subtractive Double Pass (MSDP)
imaging-spectrograph (Mein, 1977 and 1991) and Large
Coronagraph (with 530\,mm main objective) or Horizontal
Telescope (with 150\,mm main objective) installed at Bia{\l}kow
Observatory (Astronomical Institute at the University of Wroc{\l}aw)
(Radziszewski et al., 2006, 2007a, 2007b, 2008).

%Fig. 6
\begin{figure}
\centerline{\includegraphics[width=9.5cm,clip=]{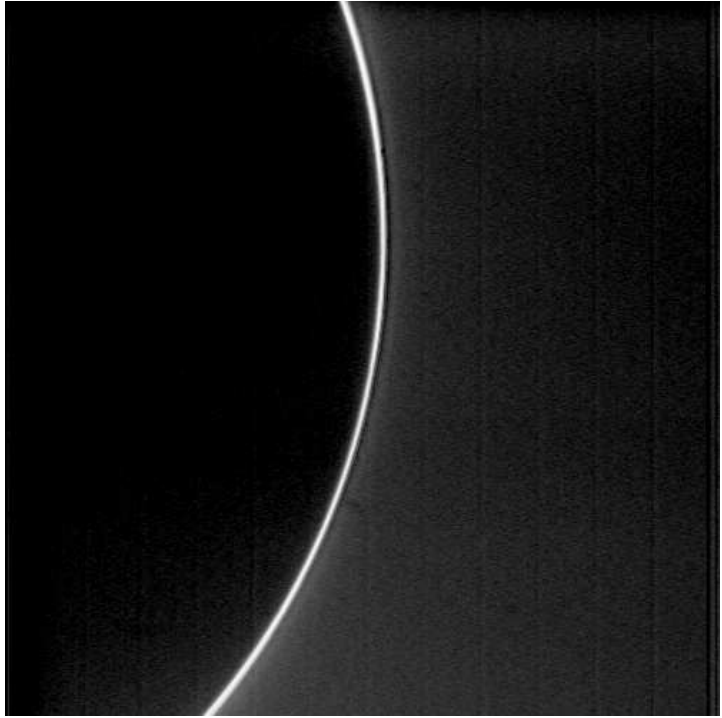}}
\caption{Image of the solar corona taken in the green line (Fe XIV 530.3\,nm)
displayed on a logarithmic intensity scale.
The profile along row 336 of the image is shown in
%Fig.~\ref{fig-8-NB-plot}. The exposure time was 65.5\,ms.
Fig.\,8. The exposure time was 65.5\,ms.
A strong radial gradient of the emission and
scattered light can be seen (compare the corresponding broad-band
%image: Fig.~\ref{fig-7-BB-image}).}
image: Fig.\,7).}
%\label{fig-6-NB-image}
\end{figure}

%Fig. 7
\begin{figure}
\centerline{\includegraphics[width=9.5cm,clip=]{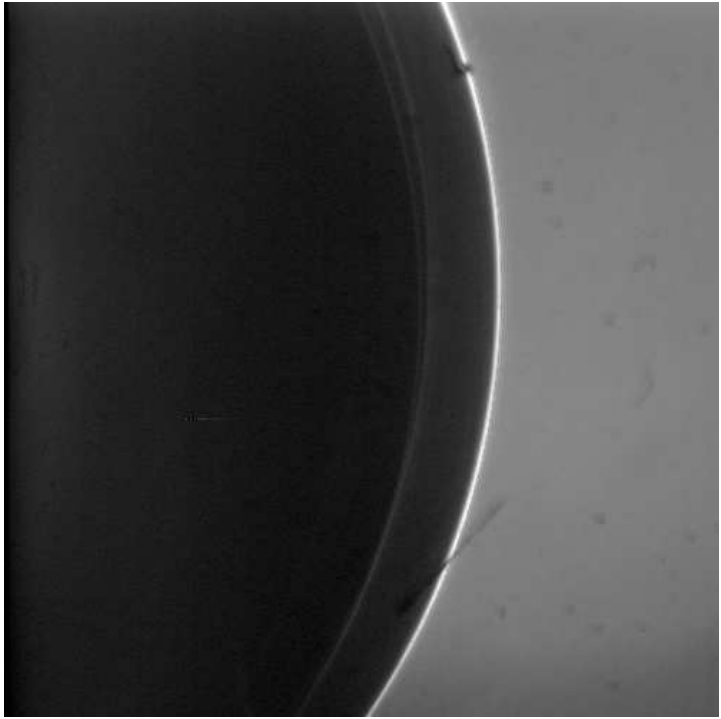}}
\caption{Image of the solar corona taken in the broad-band channel with
the broad-band and neutral density filters
displayed on a logarithmic intensity scale.
%The profile along row 300 of the image is shown in Fig.~\ref{fig-9-BB-plot}. The exposure
The profile along row 300 of the image is shown in Fig.\,9. The exposure
time was 65.5\,ms. The radial gradient of the coronal emission and the
scattered light is much weaker than in case of the narrow-band channel.
Small dust particles are apparent in this image.}
%\label{fig-7-BB-image}
\end{figure}

%Section 4
\section{Test measurements}
%\label{Tests}
Preliminary tests of the entire system were
performed in April 2009, though the lack of solar activity at that time
meant that no prominent coronal structures
were detectable.

%Subsection 4.1
\subsection{Filter tests}
%\label{Test_filter}

The filters manufactured by Barr Associates, Inc., have 50\,mm clear
diameter and have quarter-wave flatness specification. They are coated
with ion-assist refractory oxide coatings that greatly reduce the
wavelength shift with ambient temperature and filter longevity and
produce higher transmittance.

To investigate possible degradation since their last use during the
2006 eclipse, the filter passband widths and
central wavelengths were tested for thermal stability  using disk
centre solar light and a small spectrograph with a dispersion of
0.9\,nm/mm connected directly to the coronagraph (Minarovjech, 2009).
Examples of the mean reference
disk centre spectra taken with and without the narrow-band filter are
%shown in Fig.~\ref{fig-4-NB-filter-tests}. The tests showed
shown in Fig.\,4. The tests showed
that the filters have to be heated to a fairly high
temperature (45--50 degrees Celsius) in order to tune the filter
passband to the green-line wavelength.

The dependence of the main passband parameters on the filter temperature is
%displayed in Fig.~\ref{fig-5-NB-tests}.
displayed in Fig.\,5.
First, the filter passband transmission,
the width (FWHM) of its passband, and its central wavelength averaged
over the length  (7.2~mm) of the spectrograph slit were examined as a
function of filter temperature; these are indicated by the thick lines
%in Fig.~\ref{fig-5-NB-tests}. These measurements show that a typical
in Fig.\,5. These measurements show that a typical
maximum transmission of the filters is $\approx$10~\%,  and that the FWHM
of the passband is about 0.31\,nm for an ambient temperature
resulting in the filter central wavelength to equal the green line
wavelength. By taking short lengths of 0.8~mm at nine positions along
the spectrograph slit (which is aligned along the radial direction of the
filter), we also examined the variation of the same
quantities as a function of radial distance over the filter; these
%are the thin lines plotted in Fig.~\ref{fig-5-NB-tests}. There is a
are the thin lines plotted in Fig.\,5. There is a
similar dependence on
temperature in these individual measurements to the averaged results.
The filter passband transmission varies by up to about 10\,\% from the
mean value, the passband wavelength position by up to 0.02\,nm, and the
passband width by up to only 4\,\%.
The background is very stable apart from one outlying measurement.

%Subsection 4.3
\subsection{Data tests}
%\label{Data_filter}
Test observations were
taken in both the broad-band and narrow-band channels.
With solar activity at an extremely low level, no coronal structures were
visible in the green line at
that time  (2009 April 7 at 06:40 UT) but the tests were nevertheless useful in
that the optical and photometric quality of the data could be examined.
Examples of snapshots selected from the data in both
channels are
%shown in Figs.~\ref{fig-6-NB-image} and \ref{fig-7-BB-image}.
shown in Figs.\,6 and 7.
Inspection of the images shows that the image quality was very good.
Moreover, the different character of the  radial gradients illustrates the
very good spectral blocking of the narrow-band filter.
The internal
instrumental scattered light in the coronagraph and all the optical
parts of SECIS was found to be sufficiently low to
allow the required data acquisition.
This confirms
that the instrument itself is ready to measure prominent
active region coronal loops above the solar limb when they appear
under ``coronal" skies, i.e. with low degree of light scatter by
the Earth atmosphere.
%More detailed inspection of the data (Figs.~\ref{fig-8-NB-plot} and
More detailed inspection of the data (Figs.\,8 and
%\ref{fig-9-BB-plot}) shows that the noise level was low.
9) shows that the noise level was low.
%The intensity level of the data is just few digits above the dark current
%level
The photon count level in the narrow-band channel within the portion of
the image occupied by the artificial Moon was measured to be at a
very low level, averaging 8 DN s$^{-1}$, only slightly more than the dark
current level of 2--3 DN s$^{-1}$.
A bright coronal
active region is expected to have a high signal-to-noise ratio, though
experience from the 1999 and 2001 eclipses with the SECIS cameras
suggests that the cameras are very unlikely to reach saturation levels.

%Fig. 8
\begin{figure}
\centerline{\includegraphics[width=9.5cm,clip=]{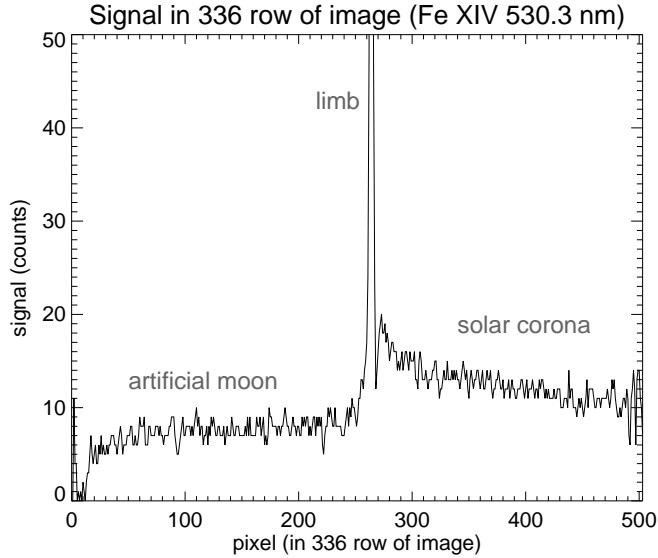}}
\caption{Plot of the signal taken from row 336 (perpendicular to
the solar limb) of the narrow-band image of the corona shown in
%Fig.~\ref{fig-6-NB-image}.
Fig.\,6.
Individual pixels covering the artificial moon,
solar limb and the solar corona are shown.}
%\label{fig-8-NB-plot}
\end{figure}

%Fig. 9
\begin{figure}
\centerline{\includegraphics[width=9.5cm,clip=]{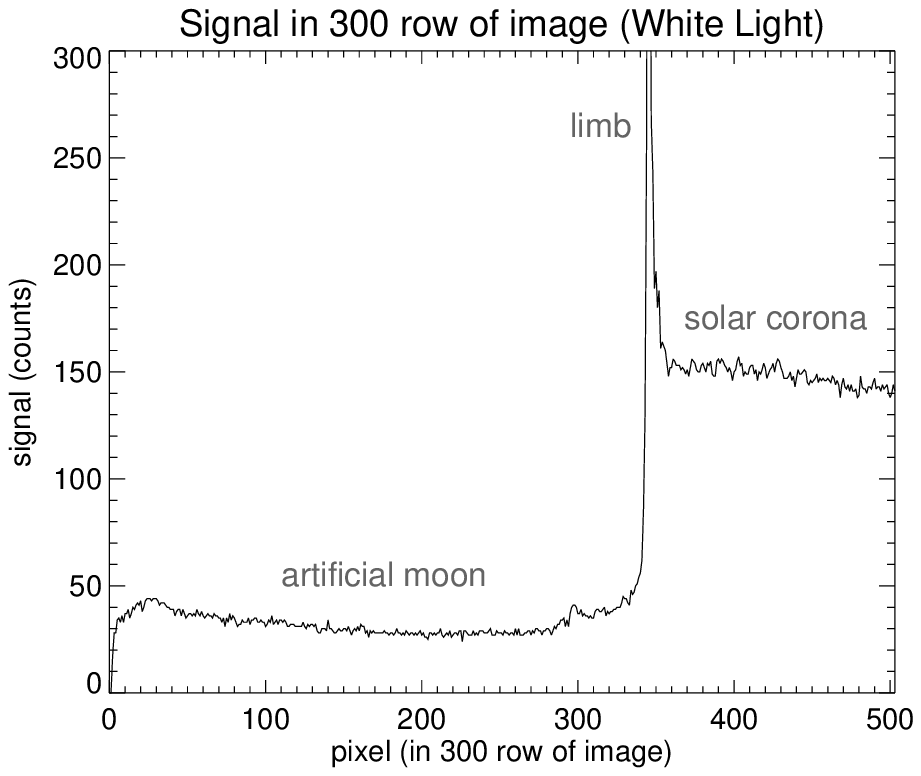}}
\caption{Plot of the signal taken from row 300
(perpendicular to the solar limb) of the narrow-band image of the corona
%shown in Fig.~\ref{fig-7-BB-image}.
shown in Fig.\,7.
Individual pixels covering the artificial moon, solar limb and the solar
corona are shown.}
%\label{fig-9-BB-plot}
\end{figure}

\section{Conclusion}
%\label{Conclusion}
The SECIS instrument installed at
the Lomnicky Peak Observatory Lyot coronagraph will allow data to
be acquired that may result in an improved knowledge of where
in the corona MHD waves are generated and/or dissipated. In
particular, the signatures of high-frequency MHD waves involved in
coronal heating may be observed. A considerable improvement in our
knowledge of a long-standing problem of solar physics could be
made by such observations, with implications for the physics of
active regions, flares, the solar wind, and solar activity, as
well as mechanisms of solar-terrestrial relationships.

%%%%%%%%%%%%%%%%%%%%%%%%%%%%%%%%%%%%%%%%%%%%%%%%%%%%%%%%%%%%%%%%%%%%%%%%%%%
% A C K N O W L E D G E M E N T S
% You must leave a blank line before the \acknowledgements command!
%%%%%%%%%%%%%%%%%%%%%%%%%%%%%%%%%%%%%%%%%%%%%%%%%%%%%%%%%%%%%%%%%%%%%%%%%%%

\acknowledgements
% Do not leave a blank line here! <---------------------->
We acknowledge the anonymous referee for comments which helped to improve the
paper.
The work of J. A. and J. R. was supported partly by the Slovak Research
and Development Agency under the contract No. APVV-0066-06 which also
covered all expenses related to the SECIS instrument at the
Lomnicky Peak Observatory (Slovakia). Authors are obliged for the
support of the Astronomical Institute, Slovak Academy of Sciences staff,
namely assistants K. Man\'{\i}k, R. Ma\v{c}ura, P. Havrila, P. Bend\'{\i}k,
and the workshop assistant J. Klein.
P.R. was supported by the Polish Ministry of Science and Higher
Education, grant number N203 022 31/2991.
This research has made use of NASA's Astrophysics Data System.

%%%%%%%%%%%%%%%%%%%%%%%%%%%%%%%%%%%%%%%%%%%%%%%%%%%%%%%%%%%%%%%%%%%%%%%%%%%%%
%                       R E F E R E N C E S
% References should start with the \begin{thebibliography}{} command,
% leaving the last curly brackets empty.
%%%%%%%%%%%%%%%%%%%%%%%%%%%%%%%%%%%%%%%%%%%%%%%%%%%%%%%%%%%%%%%%%%%%%%%%%%%%%

%%%%%%%%%%%%%%%%%%%%%%%%%%%%%%%%%%%%%%%%%%%%%%%%%%%%%%%%%%%%%%%%%%%%%%%%%%%%%
%                       H A P P Y E N D
% Your LaTeX source text must be ended by the line:
%%%%%%%%%%%%%%%%%%%%%%%%%%%%%%%%%%%%%%%%%%%%%%%%%%%%%%%%%%%%%%%%%%%%%%%%%%%%%
\end{document}